# A Martini coarse-grained model of the calcein fluorescent dye

S. Salassi[1], F. Simonelli[1], A. Bartocci[1] and G. Rossi[1*]

[1]Physics department, University of Genoa, via Dodecaneso 33, 16146, Genoa, Italy

[*]rossig@fisica.unige.it

**Abstract**

Calcein leakage assays are a standard experimental set-up to probe the extent of damage induced by external agents on synthetic lipid vesicles. The fluorescence signal associated with calcein release from liposomes is the signature of vesicle disruption, transient pore formation or vesicle fusion. This type of assay is widely used to test the membrane disruptive effect of biological macromolecules, such as proteins, antimicrobial peptides, RNA but also of synthetic nanoparticles with a polymer, metal or oxide core. Little is known about the effect that calcein, as other fluorescent dyes, may have on the properties of lipid bilayers, potentially altering their structure and permeability. Here we develop a coarse-grained model of calcein compatible with the Martini force field for lipids. We validate the model by comparing its dimerization free energy, aggregation behavior at different concentrations, and interaction with a 1-palmitoyl-2-oleoyl-*sn*-glycero-3-phosphocholine (POPC) membrane to those obtained at atomistic resolution. Our coarse-grained description of calcein makes it suitable to the simulation of large calcein-filled liposomes and of their interactions with external agents, allowing for a direct comparison between simulations and experimental liposome leakage assays.

# 1. Introduction

Calcein is a slightly water soluble, green fluorescent dye that is widely used to study cell viability and as indicator of lipid vesicle leakage by fluorescence microscopy imaging[1–3]. Calcein leakage assays are based on calcein self-quenching: at a concentration above 70 mM the fluorescence of the dye is quenched. In a typical leakage experiment, calcein is trapped into lipid vesicles or liposomes at a concentration above the self-quenching threshold. As calcein has a small permeability coefficient across a lipid bilayer[4], as long as the vesicles are intact the only fluorescence intensity that can be recorded is due to small leakage fluctuations. Varying the external conditions, such as adding an external agent to the solution, may lead to a change of the fluorescence intensity induced by calcein molecules that have leaked out from the vesicles to the surrounding environment, where calcein is dissolved at a smaller concentration. This is a signature of vesicle disruption, pore formation or vesicle fusion.

This type of assay is widely used to test the membrane disruptive effect of proteins[1], antimicrobial peptides[5], poly-cations[6], single strands of RNA[2] and also nanoparticles[7–9] (NPs). Inorganic NPs, such as metal[10,11] or oxide[12,13] NPs, are nowadays considered as multivalent agents for biomedical applications, ranging from diagnostics to therapy. Many of these applications would require some degree of control on NP-membrane interactions: once the NPs have reached their cell target, non-disruptive membrane translocation or transient membrane damage can be effectively exploited to deliver drugs to the cytoplasm, while the disruption of significant portions of the plasma membrane can be lethal to the cell.

In this perspective, liposome leakage assays, possibly performed on synthetic vesicles, are fundamental to rationalize which physico-chemical factors drive the NP towards a gentle or disruptive type of interaction with the lipid bilayer. The role played by the NP charge, for example, has been investigated using calcein leakage assays by Goodman[8] *et al.* who have found that cationic NPs induced transient poration of negatively charged liposomes, and that both cationic and anionic NPs can induce leakage from zwitterionic liposomes. Similarly, Liu[14] *et al.* has shown the transient poration due to the interaction of zwitterionic liposomes with citrate-capped Au NPs. Calcein leakage assays have been used to test the effect of varying the NP core material[15], size, shape, and surface functionalization[16] as well.

The computational approach, and the one based on molecular dynamics (MD) in particular, can certainly be a useful complement to the experiments performed on synthetic vesicles. Simulations offer the unique opportunity to track NP-membrane interactions with atomistic or molecular details. A fruitful comparison between experimental and *in silico* results, though, relies also on the possibility to simulate systems whose composition is as close as possible to that of the experimental samples. Fluorescent probes, then, come into play.

Little is known about the effect that the many different fluorescent dyes used in the experiments, sometimes at very large concentration, may have on the properties of lipid bilayers or on the interactions between them and the NPs. Calcein is not an exception: at physiological pH the molecule is expected to bring four negatively charged $COO^-$ groups, which may lead to a favorable interaction with the polar lipid head region. At large concentrations (> 100 mM) this interaction may have effects on the structural and mechanical properties of the bilayer, in turn affecting bilayer permeability and its interaction between external agents or NPs. For these reasons, it is of interest to study the calcein-membrane interactions using a computational approach.

Here we present the development of a coarse-grained (CG) model of calcein compatible with the popular Martini force field for lipids[17–19]. The model resolution makes it suitable to the simulation of calcein-filled liposomes and of their interactions with external agents.

We base our model on the structural properties of calcein, as obtained from atomistic simulations, and validate it by comparing the aggregation properties of the CG calcein in water to those obtained at atomistic level.

As calcein is charged at physiological pH, it is interesting to evaluate the performance of different versions of the Martini force field, which treat differently the electrostatic interactions. We compare three versions of Martini: the first is the standard Martini force-field[17] (STD). In STD Martini, electrostatic interactions are described as short-range interactions and the water polarizability is taken into account implicitly with a uniform dielectric constant $\varepsilon_r$ = 15. The long-range contribution to the electrostatic interactions can be added to the STD model via the particle-mesh Ewald (PME) method, an approach that has been shown[20] to improve the performance of the model when dealing with highly charged molecules. Here, we will always include the PME energy term in our STD simulations. The second model we use includes explicitly water polarizability[18] (PW). The model uses PME and a uniform dielectric constant $\varepsilon_r$ = 2.5. We also test the polarizable water model (refPW) developed by Michalowsky[19] *et al.*, in which water dielectric properties are refined with respect to the original PW model.

In the Methods section we provide all the details about the force field parameters, MD run set-up and system composition; in the Results section we present the development of the CG Martini model of calcein, and its validation in terms of dimerization free energy, aggregation behavior at different concentrations, and interaction with a 1-palmitoyl-2-oleoyl-*sn*-glycero-3-phosphocholine (POPC) membrane. Eventually, the different CG model performances and perspective used are discussed.

## 2. Methods

Both atomistic and CG simulations, as well as the analysis, are performed with the Gromacs 2016 molecular dynamics software[21].

### 2.1 Atomistic simulations

Atomistic simulations are performed with the OPLS-UA force field[22], the TIP3P water model[23] and the Berger parameters for lipids[24]. The time step is set to 2 fs. All simulations are performed in the NpT ensemble: the temperature is kept constant at 310 K by the velocity rescale thermostat[25] ($\tau_T$ = 2 ps); the pressure is kept constant at 1 bar by the Berendsen barostat ($\tau_p$ = 1 ps) for the equilibration runs and by the Parrinello-Rahman algorithm[26] ($\tau_p$ = 1 ps) for the production runs. The cut-off of the Van der Waals interactions is set to 1 nm, while the electrostatic interaction is treated via the PME method with a grid spacing of 0.12 nm.

The simulation of a single calcein in water is made in a cubic box of 5x5x5 nm$^3$ filled with water and 4 Na$^+$ counterions. The unbiased simulations with calcein at different concentrations are made in a cubic box of 10x10x10 nm$^3$ containing water, salt at physiological concentration and Na$^+$ counterions. At 30 mM calcein concentration, 20 calcein molecules are solvated by about 32000 water molecules; at 150 mM calcein concentration, 100 calcein molecules are solvated by about 30000 water molecules. The simulations for the dimerization process are made with 2 calcein molecules solvated by about 33000 water molecules and counter ions in a dodecahedron box (radius of the inscribed sphere about 5.5 nm). To study the interaction between calcein and a model zwitterionic lipid membrane we set-up a cubic box of 6x6x10 nm$^3$ with a symmetric membrane composed of 114 1-palmitoyl-2-oleoyl-*sn*-glycerol-3-phosphocholine (POPC) lipids, 7100 water molecules, salt at physiological concentration and counterions.

Simulations are initialized with the calcein molecule placed on top of the membrane, in the water phase.

## 2.2 Coarse-grained simulations

The CG simulations are performed with the Martini force-field[17]. The time step is set to 20 fs. All simulations are performed in the NpT ensemble with temperature and pressure set to 310 K and 1 bar, respectively. The velocity rescale thermostat[25] ($\tau_T$ = 1 ps), Berendsen ($\tau_p$ = 4 ps) and Parrinello-Rahman[26] ($\tau_p$ = 12 ps) barostat are used. For the simulations with the STD model the cut-off of the Van der Waals interactions is set to 1.1 nm and the dielectric constant is $\varepsilon_r$ = 15. The simulations performed with the PW model have a Van der Waals cut-off set to 1.2 nm and a dielectric constant $\varepsilon_r$ = 2.5. For the refPW model Van de Waals cut-off is 1.1 nm. In all cases, the Verlet scheme for neighbor list update and the PME method for the electrostatic interactions, with a grid spacing of 0.12 nm, are used.

The simulation boxes are set-up in the same way as in the atomistic case with the same box dimensions and number of calcein molecules. A factor 4 to 1 is considered to rescale the number of water molecules from atomistic to CG. Counterions are always present and ions at physiological concentration are present as in the atomistic case.

## 2.3 Umbrella sampling simulations

The analyses of the umbrella sampling[27] simulations are performed with the WHAM[28] algorithm using the 'gmx wham' tool[29] of the Gromacs suite. The order parameter for the umbrella sampling of the dimerization process of two calcein molecules is the distance between their centers of mass (COMs). We sample, both at atomistic and CG level, 25 windows: from a distance of 0.4 nm to 5 nm. In the atomistic case, each window is equilibrated for 10 ns followed by a production run of 50 ns. In the CG case the equilibration runs are 50 ns long while production runs are 500 ns long. The total simulated time is about 1.5 μs (atomistic) and 13.7 μs (CG).

For what concerns the umbrella sampling simulations of the desorption process of one calcein molecule from the POPC membrane we use, as order parameter, the projection along *z*-axis of the distance between the COM of the membrane and the COM of the calcein molecule. In this case 17 windows are sampled: from 1.6 nm to 4 nm; each window is initialized with a frame extracted from the corresponding unbiased simulation. At about 1.6 nm the calcein molecule is partially embedded in the membrane head region while, at distance larger than 5 nm, it is in the water phase. Each window of equilibration and production runs lasts 15 ns + 75 ns (atomistic) and 50 ns + 800 ns (CG). The total simulated time is about 1.5 μs (atomistic) and 14.5 μs (CG).

**2.4 Contacts and aggregates analysis**

The contacts analysis is performed with the 'mindist' Gromacs tool. Two atoms or beads are considered in contact if their distance is lower than a threshold of 0.8 nm. The analysis of the aggregates is performed with the 'g_aggregate' tool, a Gromacs compatible tool developed by Barnoud[30] *et al*. Two calcein molecules are assigned to the same cluster if the minimum distance between their atoms or beads is lower than a cut-off value which we set to 1.2 nm.

**3. Results and discussion**

**3.1 Atomistic model**

The parameterization of the atomistic model of calcein is based on the OPLS-UA force-field developed by Jorgensen[22] *et al.* The full details of the topology are available in our open online repository[31].

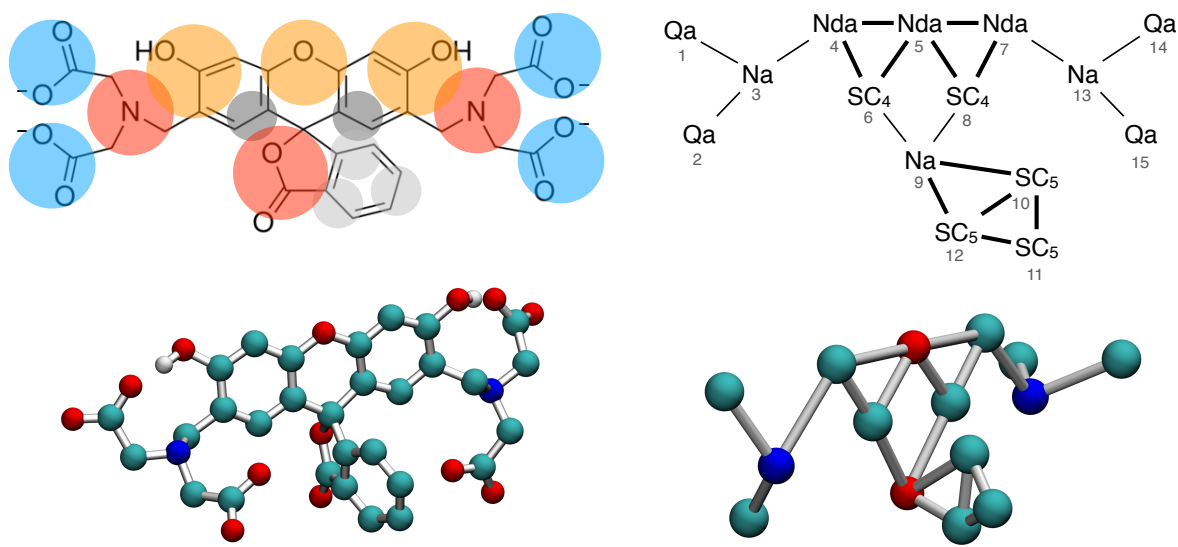

**Figure 1.** Left panel: chemical structure (top) and atomistic visualization (bottom). In the top image, colored circles indicate the mapping of atoms onto coarse grained beads. Right panel: coarse-grained description, mapping showing bead types, their numbering and the bonds between them. Thinner lines represent harmonic bonds, while thicker ones are constrained bonds. Bottom-right: visualization of a coarse-grained calcein molecule.

In the left panel of Figure 1 the chemical structure of calcein and its atomistic UA representation are shown. We have a central hydrophobic region that is highly rigid and planar and the 1-phthalanone perpendicular to the central region. The atom type assignment, the bonded and non-bonded interaction parameters, as well as the partial charge assignment, are derived from the OPLS-UA parameters for hydrocarbons and proteins[22,32,33] and from the AMBER parameters for nucleic acid and proteins[34,35]. We added improper dihedrals to impose the planarity of the four ethanoate groups (-CH$_2$-COO$^-$), that of the three rings of the central region, that of the γ-butyrolactone and that of the benzene rings. In the case of the two sp$^3$ N atoms, we added an improper dihedral in order to keep the tetrahedral geometry.

**3.2 Martini model development**

**Mapping.** In the Martini force-field, each particle (or bead) represents a chemical building block. The CG bead interactions are parameterized so as to reproduce the free energy of transfer from water to oil of the chemical group they represent. Two bead sizes are comprised within the model: normal beads have a Lennard-Jones $\sigma$ parameter of 0.47 nm and typically represent groups of 4 heavy atoms; small beads have a $\sigma = 0.43$ nm and are used to represent both groups of 3 or 2 atoms[36,37]. For the calcein molecule we adopt both normal and small bead types.

In Figure 1 the chemical structure (left panel) with the related mapping scheme and the Martini CG model (right panel) are shown. The reader is addressed to our online repository for the atomistic and CG topology files[31].

Since we are interested in the negatively charged (4$e^-$) form of the calcein molecule, all the four carboxyl groups are deprotonated and mapped onto four Qa beads[37] that carry a negative unit charge each. The other Martini types are chosen based on the logP of the chemical group they represent, or in analogy with the Martini models of other compounds, as better detailed in the Supplementary material.

We remark that the choice of the Martini bead types does not change in the STD, PW and refPW models of calcein.

**Bonded interactions.** The parameterization of the CG bonded interactions is based on the reproduction of the target atomistic distributions of distances and angles. Atomistic distributions are obtained from a trajectory of a single calcein molecule in water[38]. In Figure S1 and S2 of the Supplementary material, bond and angle distributions are shown, respectively. The equilibrium distances or angles and the force constants of the harmonic (or cosine-harmonic for angles) functions are chosen so as to reproduce the peak position and the width of the target distributions. Bond constraints are used whenever the force constant needed to reproduce the atomistic distribution is larger than 12000 kJmol$^{-1}$nm$^{-2}$. The parameterization of the benzene ring is the one proposed by De Jong[39] *et al*.

In order to keep the central region of the molecule planar, we used two improper dihedrals (between beads 4-6-5-8 and 7-8-5-6, see Figure 1) with equilibrium angles and force constants chosen to fit the atomistic distributions, as shown in Figure S3. To keep bead 9 in the same plane of the central region we added an improper dihedral between beads 5-6-8-9 and the corresponding angle distribution is shown in the left panel of Figure 2. The position of the 1-phthalanone is kept perpendicular to the central region by an improper dihedral (beads 9-12-10-11, equilibrium angle of 180° and force constant of 80 kJmol$^{-1}$rad$^{-2}$) and by an angle interaction between beads 5-9-10. At CG level, beads 9-12-10-11 are coplanar, as shown in the right panel of Figure 2.

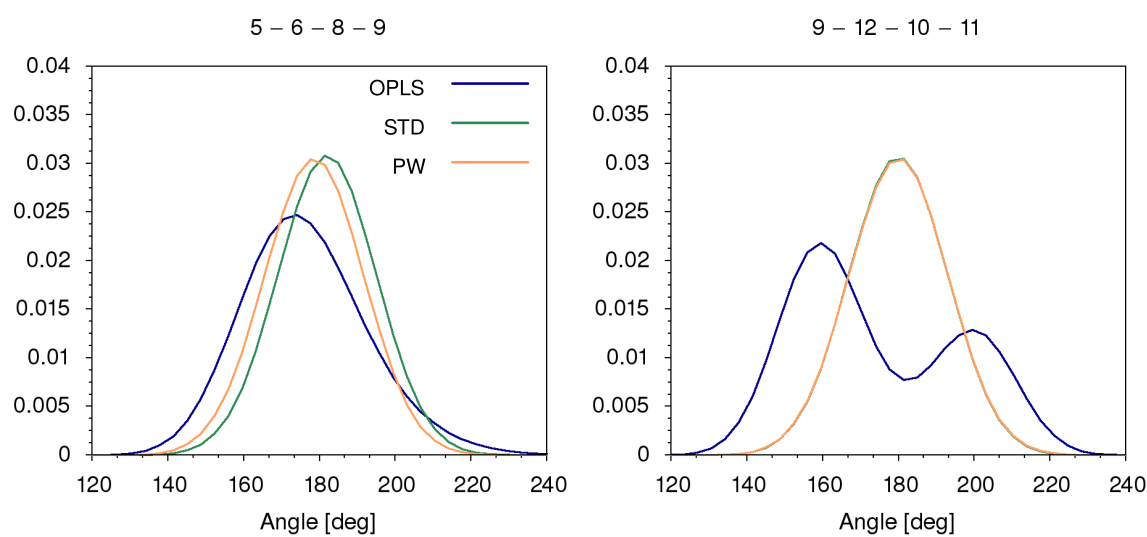

**Figure 2.** Distributions of the dihedral angle between beads 5-6-8-9 (left panel) and beads 9-12-10-11 (right panel) (See Figure 1 for beads numbering). In blue the atomistic model (OPLS), in green the coarse-grained standard Martini model (STD) and in orange the Martini model with water polarizability (PW).

## 3.3 Validation

In calcein release experiments the calcein solution inside the liposome, at high concentration (> 70 mM), is quenched by both static and dynamic quenching mechanisms[40,41]. We thus validate our CG models based on dimerization free energies and aggregation behavior of calcein at different concentrations.

**Dimerization process.** We calculate the potential of mean force (PMF) of the dimerization of two calcein molecules in water solution. The PMF is obtained from umbrella sampling simulations in which the order parameter is the distance between the COM of the two molecules. For the details about the simulation set-up see Section 2.

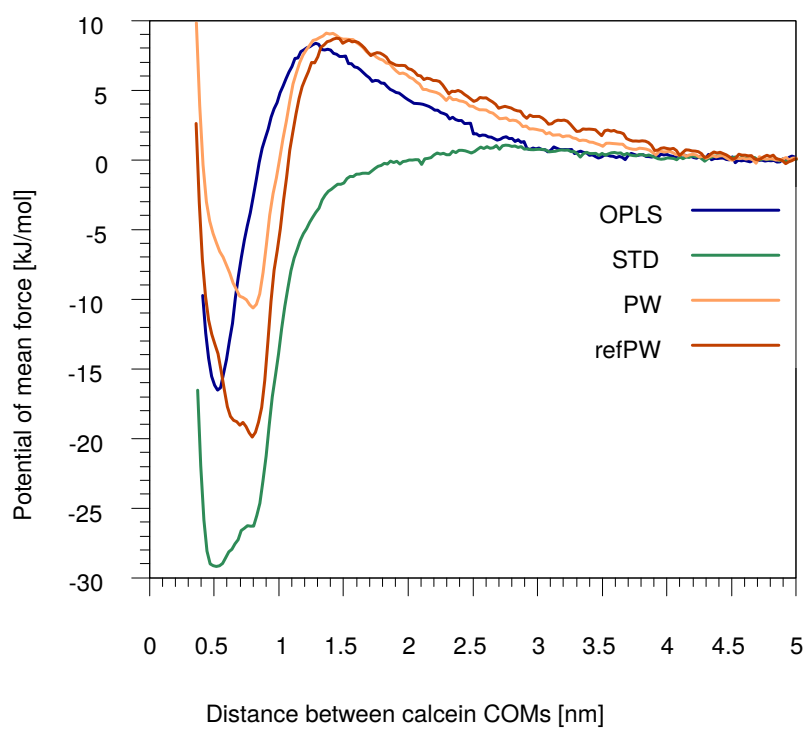

**Figure 3.** Potential of mean force (PMF) of the dimerization process of two calcein molecules in function of their distance, $d$. The PMF profiles are shifted to have a value of 0 kJ/mol in the water phase. Each profile is also shifted by the factor $2k_BT \ln(d)$ in order to take into account the Jacobian of the transformation from Cartesian to spherical coordinates[42].

A comparison of the PMF between the atomistic and CG models is shown in Figure 3. As we can see, all models agree that the bound state is more stable than the dissolved state.

The free energy difference between the unbound and bound state is about 17 kJ/mol for the atomistic model. The STD and refPW model overestimates it, with a free energy difference of ~ 29 kJ/mol and ~20 kJ/mol, respectively, while the PW model slightly underestimates it, with a difference of ~ 11 kJ/mol. The atomistic, PW and refPW models predict a small barrier for aggregation, that is not captured by the STD model. The free energy barriers for dimerization and separation are in all cases of a few $k_BT$, allowing for the dynamic formation and disruption of small aggregates on experimentally relevant time scales. The width of the dimer state well is not the same in the atomistic and CG models: the latter (especially the PW model) shows two dimer states that slightly differ in energy. In Figure 4 the two states are shown. The CG dimer that corresponds to a distance between the calcein molecules of about 0.5 nm (left panel of Figure 4, state A) is the same as that observed in the atomistic case. State A is the more populated in CG simulations with the STD model. This state is characterized by the flat contact between the central planar regions of calcein, with the 1-phthalanone ring pointing to the outer direction. The other dimer state, at about 0.8 nm (right panel of Figure 4, state B), is the one in which the benzene rings are facing each other in a sandwich π–π stacking. This state is the most populated in the CG simulations with the PW and refPW models, and is not observed in atomistic simulations.

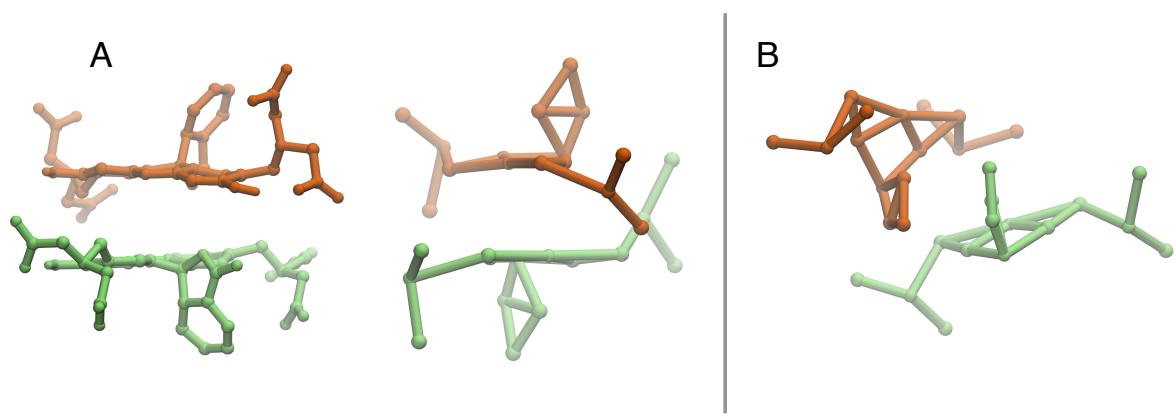

**Figure 4.** Snapshot of the dimer states. In panel A, atomistic and coarse-grained dimer state in which the distance between the calcein center of masses (COMs) is about 0.5 nm and we observe the staking of the central regions. In panel B, the coarse-grained dimer in which we observe the stacking of the benzene ring of the 1-phtalanone and the distance between COMs is about 0.8 nm.

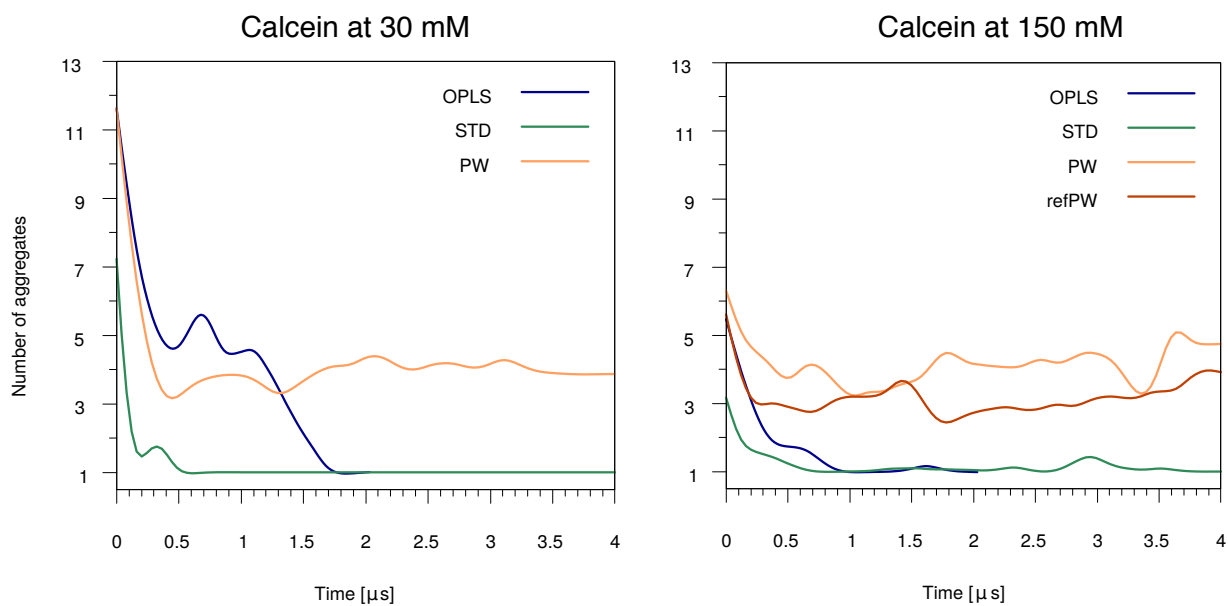

**Figure 5.** Total number of aggregates as a function of the time. Data are smoothed for clarity of visualization; the raw data are shown in Figure S4. Left panel: system with calcein at 30 mM. Right panel: system with calcein at 150 mM. In the system with calcein at 150 mM we also have tested the refined Martini polarizable water model (refPW) shown in red.

**Aggregation behavior.** The self-quenching concentration threshold for calcein is about 70 mM. We set-up atomistic and CG simulations at two different calcein concentrations, well above and well below the threshold: 30 mM and 150 mM. We use as a reference the atomistic system, and compare it to the STD and PW systems at both concentrations; for the system at 150 mM we also test the refPW. Figure 5 shows the number of aggregates as a function of time as obtained with the different models.

The atomistic simulations predict that, at both concentrations, though with different kinetics, the calcein molecules aggregate into one single cluster. The STD model agree with the atomistic result at both concentrations. The CG models with water polarizability, instead, are not able to reproduce this effect in any of our simulations. By visualizing the PW or refPW trajectories, we observe a dynamic formation-destruction of micelles of calcein molecules that do not tend to form a stable cluster. The time scales of the simulations are certainly too short to exclude the possibility of disaggregation events for the atomistic and STD models: indeed, at large calcein concentration, both the atomistic and the STD CG model predict transient disaggregation events, on time scales of microseconds.

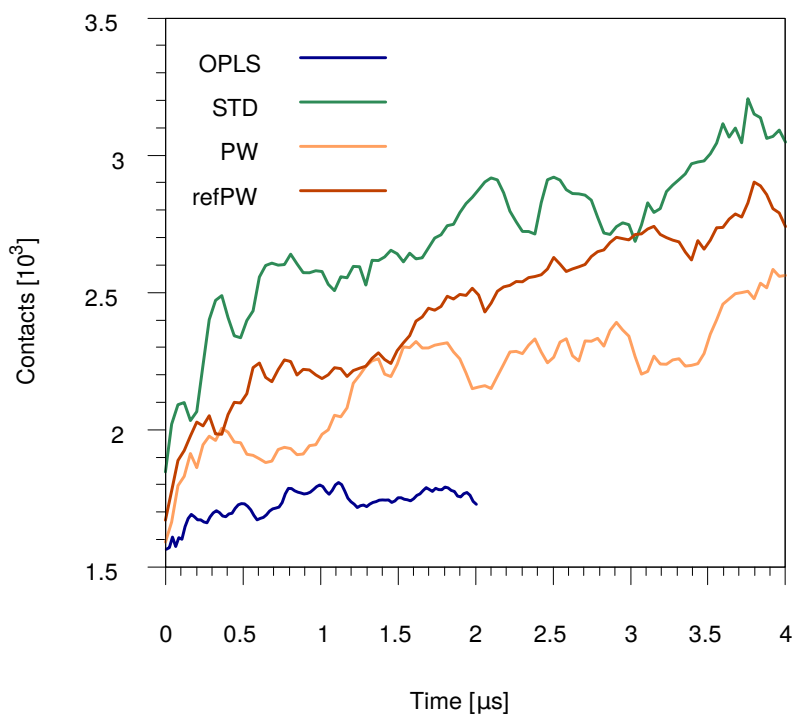

**Figure 6.** Contacts between the atoms or beads of the benzene ring of the 1-phthalanone, in the unbiased simulation of calcein at 150 mM concentration.

For what concerns the structure of the aggregates, we also observed that the PW model has the tendency to stabilize micellar aggregates in which the benzene rings of the 1-phthalanone face the interior of the micelle. On the contrary, the visual inspection of the atomistic runs suggests that most of the benzene rings are in the water phase. We calculated the number of benzene-benzene contacts, as shown in Figure 6. We see that the CG models give a number of benzene-benzene contacts larger than the atomistic model by a factor of 1.5 – 2. We conclude that the STD model is, among the CG models tested here, the one that better reproduces the atomistic aggregation behavior, characterized by the formation of large aggregates with possible detachments on time scales of microseconds;

as for the structure of the aggregates, all the CG models tested share the limitation of an overestimation of the benzene-benzene contacts.

**Calcein-membrane interactions.** In order to test the calcein-membrane interactions at atomistic and CG level, we ran a set of unbiased simulations (5.7 μs at atomistic level and 20 μs at CG level, for each model) comprising a POPC membrane (114 lipids) and a single calcein molecule, initially placed far on top of the membrane surface.

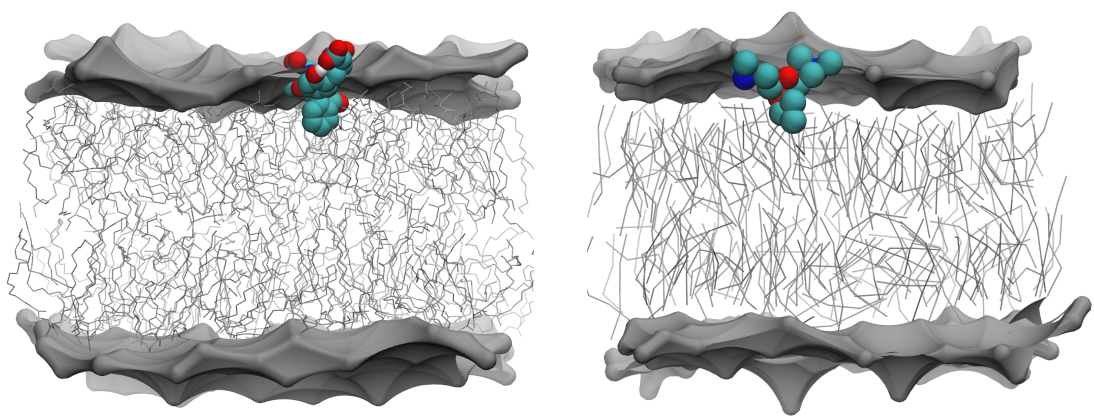

**Figure 7.** Snapshots of the membrane-bound configuration; at atomistic (left) and coarse-grained Martini standard (right) level. The head region of the lipid membrane is shown as a grey surface. The hydrophobic lipid tails are shown as thin grey sticks. The water molecules and ions, are not shown for clarity.

For the atomistic and STD models the calcein molecule binds to the membrane and unbinds spontaneously. Binding and unbinding events can be monitored by tracking the distance between calcein and the COM of phosphate groups, as shown in Figure S6 and S8. Two metastable states can be distinguished in which the calcein molecule interacts with the membrane. In the first state, or bound state, the hydrophobic benzene ring of calcein interacts with the glycerol groups of the lipids. The correlation between the calcein-COM of membrane distance and the number of contacts between the benzene ring of calcein and the lipid glycerol groups is shown in Figure S7. A snapshot of this membrane-bound configuration is shown in Figure 7 for both atomistic and STD resolutions. In CG simulations, in this configuration the calcein molecule is at a distance of about 0.1 nm from the phosphate groups; in atomistic simulations the distance is about 0.2 nm. In the second metastable state, or adsorbed state, the calcein molecule interacts with the phosphate groups only, and can be found at a distance of 1.2 nm (STD) and 0.7 nm (atomistic). The simulation with the PW model, instead, does not sample any bound or adsorbed state in the simulation time. The refPW model still predicts that the water phase is the most favorable state but also, with small probability compared to the STD and atomistic models, it predicts a bound metastable state at a distance of 0.25 nm from the phosphate group (see Figure S8).

To be more quantitative, we calculated the PMF of the binding process of one calcein from the water phase to the membrane-bound state, showed in Figure 8. The energy profile reveals two metastable states for both the atomistic and the STD model. The refPW model predicts one metastable bound state, but its most favorable configuration is in the water phase. The PMF confirms the absence of any stable state near the membrane for the PW model. The STD energy barrier to go from the adsorbed state to the bound state is about 3.5 kJ/mol, while the adsorbed state is more stable than the bound state by ~ 1 kJ/mol. The barrier for the atomistic model is higher, and it sets to 5.4 kJ/mol. The energy barrier connecting the dissolved state to the bound state for the refPW model is of 6.4 kJ/mol.

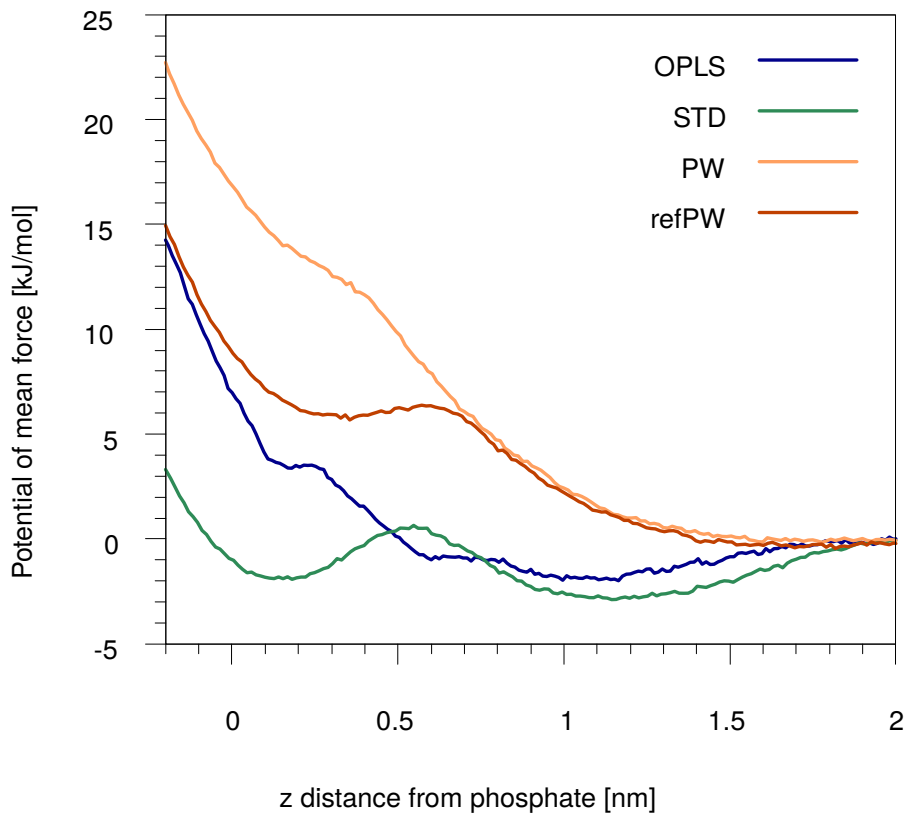

**Figure 8.** Potential of mean force (PMF) for the binding process. It is obtained from umbrella sampling simulations in which the order parameter is the *z* distance between the center of mass (COM) of the calcein molecule and the COM of the membrane. The PMF is plotted as a function of the *z* distance of the calcein COM from the phosphate groups, for clarity. The PMF profiles are shifted to zero in the water phase.

### 3.4 Discussion

We can now sum up and discuss the similarities and the differences between the three CG models of calcein we have tested. All models predict that the dimerization process is thermodynamically favored (Figure 3), in agreement with the prediction of the atomistic force field. Only the polarizable water models (PW and refPW) are able to reproduce the energy barriers of the binding and unbinding processes. The STD model, instead, overestimates the population of the dimer state over the dissolved one and does not describe the small barrier for aggregation that is instead foreseen by the PW, refPW and atomistic models. The behavior of the STD model, which we have employed here with the addition of long-range electrostatic interactions, is likely to be attributed to an overestimation of hydrophobic interactions that has been previously reported for other aromatic compounds [43,44]. A more detailed inspection of the CG dimers reveals the existence of a metastable minimum in which the benzene rings of the 1-phthalanone are in $\pi$–$\pi$ sandwich stacking, a configuration that is not sampled during the atomistic umbrella sampling calculations.

The overestimation of the hydrophobic interactions by the CG models is confirmed also by the unbiased simulations of calcein aggregation. The number of ring-ring contacts is overestimated by all CG models, and especially by the STD model. In terms of aggregation, the agreement between the PW models and the atomistic model, though, is not as satisfying as one could expect based on the dimerization free energy profile. Both the CG models that include water polarizability (PW and refPW) tend to form a highly dynamic number of calcein micelles that do not aggregate in a stable cluster. These micelles are formed by a variable number of calcein molecules, depending on the calcein concentration, as shown in Figure S5 in which the maximum size of the micelles is plotted as a function of the simulation time. A common characteristic of these micelles is that the packing of the molecules is driven by the 1-phthalanone rings that tend to stick to each other in the inner part of the micelle. At atomistic resolution, on the contrary, the packing of the molecules is driven by the stacking of the central plane of calcein. During the model optimization we explored alternative mappings with the aim to improve the performance of the PW models in this respect. We have tried to have a less hydrophobic benzene with the introduction of a new Martini bead type that is equal to the SC5 bead but has a decreased self-interaction (− 6 %) and an increased interaction with the water beads (+ 6.5 %). The improvement is little and does not justify the introduction of a new bead type. The final PW model uses an Nda type to describe the central C-O-C group of calcein. A straightforward choice based on standard Martini mappings would suggest to use the Na type, which has only hydrogen acceptor character. The Nda-Nda interactions, though, are stronger than the Na-Na interactions (4.5 kJ/mol vs.

4 kJ/mol), and thus provide a better aggregation behavior. We have tried also to increase further the self-interaction of the beads of the central plane (beads 4, 5 and 7, up to +11 %). Again, this fine tuning did not significantly improve the aggregation behavior of the PW or refPW models. Overall and especially in terms of aggregation at high concentration, the behavior of the STD (with long-range electrostatics) model seems in qualitative agreement with the atomistic one.

Experimentally, fluorescence-lifetime measurements[41] suggest that calcein's self-quenching method is based on both static quenching and dynamic quenching, the latter prevailing over the former. The results obtained with the atomistic and CG STD model would suggest that the static quenching deriving from calcein-calcein aggregation (over time scales longer than the fluorescence lifetime, 4 ns) is the main mechanism of fluorescence quenching; PW models, instead, show an aggregation behavior which is more consistent with a dynamic quenching mechanism.

Our simulations show a good agreement between the STD model predictions and the atomistic ones in terms of interaction with the lipid membrane. Both the STD and atomistic models indicate the existence of two metastable states in which the calcein is in close contact with the membrane (the bound state is shown in Figure 7). The PW model, on the contrary, shows no interaction at all between the negatively charged calcein and the membrane surface (Figure 8). The refPW model slightly improves the situation, but it does not predict stable binding either (Figure 8). This latter behavior is at odds with the general experimental observation of stable interactions between phosphatidylcholine membranes and anionic macromolecules[45] and nanoparticles[9,46]. A possible explanation of this behavior could be the much larger hydration of the lipid head groups that is provided by the PW model. We calculated the density of water as a function of the $z$ component of the distance from the COM of a POPC membrane, in absence of calcein, with the OPLS-UA,

STD, PW and refPW models. Table S1 contains the water density recorded in the lipid phosphate plane, and shows that the STD underestimates the atomistic water density by 4%, while PW and refPW overestimate it by 11% and 14%, respectively. Head group hydration could screen the favorable interaction of the negatively charged groups of calcein with the membrane choline groups. These observations lead us to the conclusion that the use of the STD model of calcein could be preferred over the use of the PW model, especially when looking at calcein in presence of a lipid membrane.

We anticipate that our CG model of calcein will be useful to the interpretation of vesicle and liposome leakage assays. Different external agents, such as cell penetrating peptides, polymer and inorganic NPs, can damage the lipid bilayer via different mechanisms and to a different extent. In order to achieve a molecular interpretation of the experimental data, the possibility to include the leaking dye in the simulations may help to clarify the role played by the dye itself during membrane poration.

Calcein is used also as model metabolite to study the permeability of gap junction channels. Zonta[47] *et al.,* for example, used calcein to probe the permeability of the human connexin26 channel protein to negatively charged molecules bearing different total charge. They found that while the neutral form of calcein is able to cross the channel and the transition rate is comparable to that obtained experimentally, the channel is impermeable to the fully deprotonated (-4$e^-$) calcein. With our CG model it would be relatively easy to change the protonation state of the dye and assess its role during translocation[48], accessing longer time scales than with a pure atomistic approach.

**4. Conclusions**

In this paper we have developed and tested a coarse-grained model of calcein, a fluorescent dye routinely used in liposome leakage assays. The model is compatible with the Martini force field, and as such is suitable to the simulation, via molecular dynamics, of the interaction between synthetic nanoparticles and large lipid bilayers (e.g. calcein-containing liposomes with diameters of tens of nm). The availability of a coarse-grained model of calcein is a step forward in the direction of a better integration of experimental and computational data, as the model allows to perform simulations in which the effect of the dye during the membrane poration process is taken into account explicitly and with molecular resolution.

**Acknowledgements**

Giulia Rossi acknowledges funding from the ERC Starting Grant BioMNP – 677513. Part of the calculations are performed at CINECA under the HP10CL5BTC grant.

# References

[1] Allen T M and Cleland L G 1980 Serum-induced leakage of liposome contents *BBA - Biomembr.* **597** 418–26

[2] Tsao Y S and Huang L 1985 Sendai Virus Induced Leakage of Liposomes Containing Gangliosides *Biochemistry* **24** 1092–8

[3] Maherani B, Arab-Tehrany E, Kheirolomoom A, Geny D and Linder M 2013 Calcein release behavior from liposomal bilayer; Influence of physicochemical/mechanical/structural properties of lipids *Biochimie* **95** 2018–33

[4] Shimanouchi T, Ishii H, Yoshimoto N, Umakoshi H and Kuboi R 2009 Calcein permeation across phosphatidylcholine bilayer membrane: Effects of membrane fluidity, liposome size, and immobilization *Colloids Surfaces B Biointerfaces* **73** 156–60

[5] Zhang L, Rozek A and Hancock R E W 2001 Interaction of Cationic Antimicrobial Peptides with Model Membranes *J. Biol. Chem.* **276** 35714–22

[6] Kepczynski M, Jamróz D, Wytrwal M, Bednar J, Rzad E and Nowakowska M 2012 Interactions of a hydrophobically modified polycation with zwitterionic lipid membranes *Langmuir* **28** 676–88

[7] Verma A, Uzun O, Hu Y, Hu Y, Han H-S, Watson N, Chen S, Irvine D J and Stellacci F 2008 Surface-structure-regulated cell-membrane penetration by monolayer-protected nanoparticles *Nat. Mater.* **7** 588–95

[8] Goodman C M, McCusker C D, Yilmaz T and Rotello V M 2004 Toxicity of Gold Nanoparticles Functionalized with Cationic and Anionic Side Chains *Bioconjug. Chem.* **15** 897–900

[9] Van Lehn R C, Atukorale P U, Carney R P, Yang Y S, Stellacci F, Irvine D J and Alexander-Katz A 2013 Effect of particle diameter and surface composition on the

spontaneous fusion of monolayer-protected gold nanoparticles with lipid bilayers *Nano Lett.* **13** 4060–7

[10]   Dreaden E C, Alkilany A M, Huang X, Murphy C J and El-Sayed M A 2012 The golden age: gold nanoparticles for biomedicine *Chem. Soc. Rev.* **41** 2740–79

[11]   Yang Y S, Carney R P, Stellacci F and Irvine D J 2014 Enhancing radiotherapy by lipid nanocapsule-mediated delivery of amphiphilic gold nanoparticles to intracellular membranes *ACS Nano* **8** 8992–9002

[12]   Dumontel B, Canta M, Engelke H, Chiodoni A, Racca L, Ancona A, Limongi T, Canavese G and Cauda V 2017 Enhanced biostability and cellular uptake of zinc oxide nanocrystals shielded with a phospholipid bilayer *J. Mater. Chem. B* **5** 8799–813

[13]   Riedinger A, Guardia P, Curcio A, Garcia M A, Cingolani R, Manna L and Pellegrino T 2013 Subnanometer local temperature probing and remotely controlled drug release based on Azo-functionalized iron oxide nanoparticles *Nano Lett.* **13** 2399–406

[14]   Wang F and Liu J 2015 Self-healable and reversible liposome leakage by citrate-capped gold nanoparticles: probing the initial adsorption/desorption induced lipid phase transition *Nanoscale* **7** 15599–604

[15]   Liu J 2016 Interfacing zwitterionic liposomes with inorganic nanomaterials: Surface forces, membrane integrity, and applications *Langmuir* **32** 4393–404

[16]   Carney R P, Carney T M, Mueller M and Stellacci F 2012 Dynamic cellular uptake of mixed-monolayer protected nanoparticles. *Biointerphases* **7** 17

[17]   Marrink S J, Risselada H J, Yefimov S, Tieleman D P and de Vries A H 2007 The MARTINI Force Field: Coarse Grained Model for Biomolecular Simulations *J. Phys. Chem. B* **111** 7812–24

[18]   Yesylevskyy S O, Schäfer L V., Sengupta D and Marrink S J 2010 Polarizable Water Model for the Coarse-Grained MARTINI Force Field ed M Levitt *PLoS Comput. Biol.* **6** e1000810

[19]   Michalowsky J, Schäfer L V., Holm C and Smiatek J 2017 A refined polarizable water model for the coarse-grained MARTINI force field with long-range electrostatic interactions *J. Chem. Phys.* **146** 54501

[20]   Lee H and Larson R G 2009 Multiscale modeling of dendrimers and their interactions with bilayers and polyelectrolytes. *Molecules* **14** 423–38

[21]   Abraham M J, Murtola T, Schulz R, Páll S, Smith J C, Hess B and Lindah E 2015 Gromacs: High performance molecular simulations through multi-level parallelism from laptops to supercomputers *SoftwareX* **1–2** 19–25

[22]   Jorgensen W L, Madura J D and Swenson C J 1984 Optimized Intermolecular Potential Functions for Liquid Hydrocarbons *J. Am. Chem. Soc.* **106** 6638–46

[23]   Jorgensen W L, Chandrasekhar J, Madura J D, Impey R W and Klein M L 1983 Comparison of simple potential functions for simulating liquid water *J. Chem. Phys.* **79** 926–35

[24]   Berger O, Edholm O and Jähnig F 1997 Molecular dynamics simulations of a fluid bilayer of dipalmitoylphosphatidylcholine at full hydration, constant pressure, and constant temperature *Biophys. J.* **72** 2002–13

[25]   Bussi G, Donadio D and Parrinello M 2007 Canonical sampling through velocity rescaling *J. Chem. Phys.* **126**

[26]   Parrinello M and Rahman A 1981 Polymorphic transitions in single crystals: A new molecular dynamics method *J. Appl. Phys.* **52** 7182–90

[27]   Kästner J 2011 Umbrella sampling *Wiley Interdiscip. Rev. Comput. Mol. Sci.* **1** 932–42

[28] Kumar S, Rosenberg J M, Bouzida D, Swendsen R H and Kollman P A 1992 THE weighted histogram analysis method for free-energy calculations on biomolecules. I. The method *J. Comput. Chem.* **13** 1011–21

[29] Hub J S, de Groot B L and van der Spoel D 2010 g_wham—A Free Weighted Histogram Analysis Implementation Including Robust Error and Autocorrelation Estimates *J. Chem. Theory Comput.* **6** 3713–20

[30] Castillo N, Monticelli L, Barnoud J and Tieleman D P 2013 Free energy of WALP23 dimer association in DMPC, DPPC, and DOPC bilayers *Chem. Phys. Lipids* **169** 95–105

[31] https://bitbucket.org/biomembnp/biomembnp

[32] Jorgensen W L and Tirado-Rives J 1988 The OPLS Potential Functions for Proteins. Energy Minimizations for Crystals of Cyclic Peptides and Crambin *J. Am. Chem. Soc.* **110** 1657–66

[33] Briggs J M, Nguyen T B and Jorgensen W L 1991 Monte Carlo simulations of liquid acetic acid and methyl acetate with the OPLS potential functions *J. Phys. Chem.* **95** 3315–22

[34] Weiner S J, Kollman P A, Case D A, Singh U C, Ghio C, Alagona G G, Profeta S, Weinerl P and Weiner P 1984 A New Force Field for Molecular Mechanical Simulation of Nucleic Acids and Proteins *J . Am. Chem. SOC* **106** 765–84

[35] Cornell W D, Cieplak P, Bayly C I, Gould I R, Merz K M, Ferguson D M, Spellmeyer D C, Fox T, Caldwell J W and Kollman P A 1995 A Second Generation Force Field for the Simulation of Proteins, Nucleic Acids, and Organic Molecules *J. Am. Chem. Soc.* **117** 5179–97

[36] Lee H, De Vries A H, Marrink S J and Pastor R W 2009 A coarse-grained model for polyethylene oxide and polyethylene glycol: Conformation and hydrodynamics *J.*

*Phys. Chem. B* **113** 13186–94

[37]   Monticelli L, Kandasamy S K, Periole X, Larson R G, Tieleman D P and Marrink S J 2008 The MARTINI coarse-grained force field: Extension to proteins *J. Chem. Theory Comput.* **4** 819–34

[38]   Wassenaar T A, Pluhackova K, Böckmann R A, Marrink S J and Tieleman D P 2014 Going backward: A flexible geometric approach to reverse transformation from coarse grained to atomistic models *J. Chem. Theory Comput.* **10** 676–90

[39]   De Jong D H, Singh G, Bennett W F D, Arnarez C, Wassenaar T A, Schäfer L V., Periole X, Tieleman D P and Marrink S J 2013 Improved parameters for the martini coarse-grained protein force field *J. Chem. Theory Comput.* **9** 687–97

[40]   Andersson A, Danielsson J, Gräslund A and Mäler L 2007 Kinetic models for peptide-induced leakage from vesicles and cells *Eur. Biophys. J.* **36** 621–35

[41]   Patel H, Tscheka C and Heerklotz H 2009 Characterizing vesicle leakage by fluorescence lifetime measurements *Soft Matter* **5** 2849

[42]   Trzesniak D, Kunz A-P E and van Gunsteren W F 2007 A Comparison of Methods to Compute the Potential of Mean Force *ChemPhysChem* **8** 162–9

[43]   Bochicchio D, Panizon E, Ferrando R, Monticelli L and Rossi G 2015 Calculating the free energy of transfer of small solutes into a model lipid membrane: Comparison between metadynamics and umbrella sampling *J. Chem. Phys.* **143** 0–7

[44]   Bereau T and Kremer K 2015 Automated Parametrization of the Coarse-Grained Martini Force Field for Small Organic Molecules *J. Chem. Theory Comput.* **11** 2783–91

[45]   Michanek A, Kristen N, Höök F, Nylander T and Sparr E 2010 RNA and DNA interactions with zwitterionic and charged lipid membranes - A DSC and QCM-D study *Biochim. Biophys. Acta - Biomembr.* **1798** 829–38

[46]   Wang B, Zhang L, Chul S and Granick S 2008 Nanoparticle-induced surface reconstruction of phospholipid membranes *PNAS* **105** 18171

[47]   Zonta F, Polles G, Sanasi M F, Bortolozzi M and Mammano F 2013 The 3.5 ångström X-ray structure of the human connexin26 gap junction channel is unlikely that of a fully open channel *Cell Commun. Signal.* **11** 1

[48]   Canepa E, Salassi S, Simonelli F, Ferrando R, Rolandi R, Lambruschini C, Canepa F, Relini A and Rossi G 2018 Nanoparticle-lipid membrane interactions: the protonation of anionic nanoparticles at the membrane surface reduces membrane disruption *Submitted*